\documentclass[sigplan,10pt]{acmart}

\renewcommand\footnotetextcopyrightpermission[1]{} 
\acmConference[ATC '26]{2026 ACM SIGOPS Annual Technical Conference}{November 15--18, 2026}{Hong Kong}
\acmYear{2026}

\setcopyright{none}             
\usepackage{nicefrac}
\usepackage{siunitx}
\usepackage{array,framed}
\usepackage{booktabs}
\usepackage{color, float, epsfig, wrapfig, graphics, graphicx, subcaption}
\usepackage{textcomp}
\usepackage{setspace}
\usepackage{latexsym,fancyhdr,url}
\usepackage{enumerate}
\usepackage[ruled,vlined]{algorithm2e} 
\usepackage{algpseudocode}
\usepackage{xparse}
\usepackage{xspace}
\usepackage{multirow}
\usepackage{csvsimple}
\usepackage{balance} 
\usepackage{tikz}
\usepackage{xurl}
\usepackage{booktabs}
\usepackage{multirow}
\usepackage{tabularx}
\usepackage{booktabs}
\usepackage{makecell}
\usepackage{microtype}
\usepackage{tabularx}

\usepackage[dvipsnames]{xcolor}
\usetikzlibrary{shapes.geometric, arrows.meta, positioning, fit, backgrounds, calc, shadows}

\usepackage{tikz, pgfplots, pgfplotstable}
\usetikzlibrary{shapes.geometric, arrows, external, pgfplots.groupplots, matrix}
\pgfplotsset{compat=1.9}

\usepackage{mathtools}

\DeclareMathAlphabet{\mathcal}{OMS}{cmsy}{m}{n}

\DeclareGraphicsExtensions{.png,.PNG,.pdf,.PDF,.jpg,.mps,.jpeg}

\newcommand{\bheading}[1]{{\vspace{2pt}\noindent{\textbf{#1}}}}

\newcounter{note}[section]

\newcommand{\yz}[1]{}
\newcommand{\qz}[1]{}
\newcommand{\ky}[1]{}

\newcommand{\sysname}{\textsc{IntCC}\xspace}

\newcommand{\secref}[1]{\mbox{Sec.~\ref{#1}}\xspace}

\newcommand{\figref}[1]{\mbox{Figure~\ref{#1}}}
\newcommand{\tabref}[1]{\mbox{Table~\ref{#1}}}

\newcommand{\ignore}[1]{}

\newcommand{\eg}{\textit{e.g.}\xspace}

\newcounter{packednmbr}

\newenvironment{packeditemize}{
\begin{list}{$\bullet$}{
\setlength{\labelwidth}{0pt}
\setlength{\itemsep}{2pt}
\setlength{\leftmargin}{\labelwidth}
\addtolength{\leftmargin}{\labelsep}
\setlength{\parindent}{0pt}
\setlength{\listparindent}{\parindent}
\setlength{\parsep}{1pt}
\setlength{\topsep}{1pt}}}{\end{list}}

\begin{document}
\pagestyle{plain} 
\title{\sysname: A Framework for Interactive Confidential Computing}


\author{Qingzhe Bing}
\authornote{Both authors contributed equally to this research.}
\affiliation{%
  \institution{SUSTech}
  \city{Shenzhen}
  \country{China}}
\email{12432651@mail.sustech.edu.cn}

\author{Kaiyuan Zhang}
\authornotemark[1]
\affiliation{%
  \institution{SUSTech}
  \city{Shenzhen}
  \country{China}}
\email{12432720@mail.sustech.edu.cn}

\author{Yinqian Zhang}
\correspondingauthor
\affiliation{%
  \institution{SUSTech}
  \city{Shenzhen}
  \country{China}}
\email{yinqianz@acm.org}

\begin{abstract}

Confidential computing leverages Trusted Execution Environments (TEEs) to ensure the confidentiality and integrity of data in use. However, TEEs rely on remote attestation to guarantee the integrity of their initial memory state. This model is fundamentally at odds with interactive development workflows. In scenarios like LLM fine-tuning and exploratory data analysis, data processors need human-in-the-loop capabilities, including dynamic code injection, intermediate state inspection, and hyperparameter tuning, all of which inherently violate the static, one-time integrity guarantees of traditional remote attestation. 

To reconcile this tension, we propose the interactive confidential computing paradigm, a system architecture enabling untrusted data processors to execute dynamic, non-deterministic operations within TEEs without compromising data confidentiality. Driven by the insight that inherently unmeasurable human interaction must be excluded from the Trusted Computing Base (TCB), we logically partition the TEE into an interactive controller and a  verifiable runtime. To realize this paradigm, we present \sysname, a framework featuring three key mechanisms: (1) a proxy-based dispatch system to preserve the native development experience; (2) a fine-grained information flow control mechanism based on a security lattice to prevent data leakage; and (3) a privacy-preserving verifiable execution mechanism to guarantee the runtime compliance of dynamic workflows. We implement \sysname on AMD SEV-SNP using Confidential Containers and evaluate it across diverse real-world workloads. Our experiments demonstrate that \sysname effectively balances security and interactivity, incurring a practical overhead of less than 5\% for LLM fine-tuning and under 17\% for data analysis relative to baseline execution.
\end{abstract}



\maketitle


\section{Introduction}

The confidential computing paradigm is reshaping cloud security by ensuring that data remains protected during processing~\cite{bues2025unlocking}. At its core, this paradigm relies on hardware-based Trusted Execution Environments (TEEs) to establish secure execution contexts through strict memory isolation and transparent hardware-level encryption. Modern TEEs provide protection at various granularities, ranging from process-level enclaves (\eg, Intel SGX~\cite{intel_sgx}) to full-system Confidential Virtual Machines (CVMs) (\eg, AMD SEV-SNP~\cite{amd_sev} and Intel TDX~\cite{intel_tdx}). By shielding the execution state from the underlying untrusted host operating system or hypervisor, these technologies promise to protect sensitive assets from cloud providers, effectively enabling the secure deployment of cloud-native workloads~\cite{coco_website, ohrimenko2016oblivious}.

However, a fundamental paradox lies at the heart of confidential computing: the rigid, static security model of TEEs is inherently incompatible with the dynamic nature of modern interactive workloads. TEEs rely on remote attestation to guarantee the integrity of their initial memory state~\cite{RATS}. This mechanism binds the trust of the system to a static measurement of the code and data loaded at boot time. This model is fundamentally at odds with interactive workflows. In scenarios like Large Language Model (LLM) fine-tuning~\cite{peft,llamafactory,peng2023gpt4llm} or Explorato\-ry Data Analysis (EDA), data processors require human-in-the-loop capabilities, including dynamic code injection, intermediate state inspection, and hyperparameter tuning. These operations are inherently dynamic and cannot be anticipated at launch time, placing them beyond the coverage of traditional remote attestation.

This conflict is explicitly acknowledged by leading industry frameworks. For instance, the Confidential Containers (CoCo) architecture~\cite{coco_website} warns that enabling interactive features like \texttt{kubectl exec} fundamentally disrupts the static chain of trust established by remote attestation; consequently, such tools are strictly disabled in production profiles~\cite{ConfidentialcontainersReleasesV0170md}. Red Hat practitioners similarly highlight this dilemma, noting that while diagnosing production issues is crucial, enabling standard debugging ``risks exposing sensitive data to the entities we aim to exclude from the trust model, undermining the core principles of confidential computing''~\cite{HowDebugConfidential}. Furthermore, the Confidential Computing Consortium warns that unrestricted debugging interfaces render it ``almost impossible to ensure the protection of the confidentiality and integrity of the workload''~\cite{consortiumConfidentialComputingLogging}.

Crucially, while existing industry guidelines primarily focus on mitigating threats from untrusted cloud providers, interactive workloads exacerbate the inherent mutual distrust between data owners and data processors. On one hand, data owners cannot safely provision interactive access: once post-launch operations are permitted, a data processor could pivot from a benign tenant executing a pre-measured workload into an active adversary injecting malicious logic to exfiltrate private assets. On the other hand, data processors are reluctant to trust data owners or external auditors. Because interactive workflows embed proprietary algorithms, hyperparameters, and custom logic, processors are fundamentally unwilling to expose their source code or plaintext execution traces for compliance auditing. This dichotomy leaves a critical research gap: how can we enable untrusted data processors to execute non-deterministic, human-in-the-loop workflows on sensitive data, while guaranteeing verifiable runtime compliance for the data owner without compromising the processor's intellectual property?

To address this problem, we present \textit{interactive confidential computing}, a paradigm that enables untrusted entities to perform dynamic, non-deterministic operations within TEE-protected boundaries while still providing guarantees for data confidentiality and verifiable runtime compliance. The key insight is that because human interaction is inherently arbitrary and unmeasurable, they must be excluded from the Trusted Computing Base (TCB). The underlying architecture for interactivity must therefore be a logical partition: the unmeasurable interactive logic is stripped from the TCB and isolated into an \textit{Interactive Controller}, while measurable sensitive operations are confined within a \textit{Verifiable Runtime}. Crucially, the fundamental contract bridging these two partitions is a prescribed \textit{security policy}. Under this paradigm, data processors can utilize familiar ecosystems (\eg, PyTorch~\cite{pytorch}) seamlessly; however, all cross-boundary interactions and operations on sensitive assets are strictly mediated by the Verifiable Runtime to ensure absolute compliance with the agreed-upon security policy. 

To realize this partitioned architecture, we propose the \sysname framework to address the critical challenges of maintaining confidentiality and ensuring integrity under dynamic interaction. First, to guarantee performance while preserving the native development experience, we design a proxy-based dispatch system. This system maintains lightweight shadow objects in the controller that reference concrete data within the runtime, enabling data processors to operate on remote assets using standard APIs without incurring the prohibitive overhead of cross-domain serialization. Second, simply separating the domains is insufficient if the interactive controller can infer sensitive data. We solve this by designing a fine-grained information flow control mechanism~\cite{denningifc, sabelfeld2003language} based on a security lattice that enforces relaxed noninterference, ensuring that only data adhering to declassification policies can cross the security boundary. Third, to establish runtime compliance for dynamically determined execution paths without exposing the data processor's proprietary code to data owners or auditors, we introduce a \textit{privacy-preserving} verifiable execution mechanism. This mechanism generates canonical audit evidence for dynamic operations, commits to the execution trace, and proves in zero-knowledge that the committed trace satisfies predefined policies.

We instantiate \sysname on AMD SEV-SNP~\cite{amd_sev} based Kata Containers~\cite{kata} and evaluate it across real-world interactive workloads, including LLM fine-tuning and data analysis. The results demonstrate that \sysname supports these workflows with practical overhead relative to the baseline, while effectively mitigating data leakage risks~\cite{zhu2019deep, geiping2020inverting, fredrikson2015model}.

In summary, the contributions are as follows:
\begin{packeditemize}

    \item We introduce the \textbf{interactive confidential computing} paradigm to reconcile the fundamental tension between the static integrity guarantees of TEE remote attestation and the inherent dynamism of interactive development.
    
    \item We architect the \textbf{\sysname} framework to operationalize this paradigm. \sysname integrates three core mechanisms: (1) a proxy-based dispatch system to preserve native usability and performance; (2) a fine-grained information flow control mechanism to strictly enforce data confidentiality; and (3) a privacy-preserving verifiable execution mechanism to guarantee runtime compliance.

    
    \item We implement \sysname on AMD SEV-SNP and evaluate it against real-world workloads. Our results demonstrate that \sysname effectively thwarts data exfiltration while incurring highly practical overhead (e.g., $<5\%$ for LLM fine-tuning and $<17\%$ for data analysis).

\end{packeditemize}

\section{Background}

\subsection{Confidential Computing}

Confidential computing leverages TEEs to protect data in use. Modern CVMs, such as AMD SEV-SNP~\cite{amd_sev} and Intel TDX~\cite{intel_tdx}, incorporate the Guest OS and application stack into the TCB. This architectural design substantially reduces migration barriers for complex workloads like deep learning~\cite{pytorch, abadi2016tensorflow}. Furthermore, to accommodate cloud-native deployments, projects like CoCo seamlessly encapsulate these unmodified workloads within CVMs~\cite{kata, coco_website}.

Remote attestation serves as the foundational trust primitive in confidential computing. It enables remote verifiers to validate that a CVM initial state, encompassing firmware, kernel, and loaded images, conforms to expected values via measurement reports~\cite{RATS}. However, this rigid security model exposes a fundamental coverage gap for interactive workloads: it cannot attest to non-deterministic, post-launch interactions introduced by human operators. Consequently, to preserve strict security assurances, existing frameworks enforce a black-box model that severely restricts external interactions~\cite{ConfidentialcontainersReleasesV0170md}. While maximizing security, this inherently conflicts with the dynamic, iterative nature and observability requirements of modern cloud-native applications.

\subsection{Application Scenarios for Interactive Confidential Computing}

While the black-box model of TEEs is effective for static tasks, production environments frequently demand interactive workflows where computational logic cannot be fully determined prior to launch. We identify three representative scenarios driving this need:

\begin{packeditemize}
\item \textbf{LLM Fine-tuning:} LLM fine-tuning~\cite{peft, llamafactory,hu2022lora} is inherently iterative and human-in-the-loop. Data processors typically cannot determine the optimal configuration in a single pass; they require runtime monitoring during training. For instance, they must observe the loss function curve to detect overfitting, dynamically adjust hyperparameters such as learning rate or batch size based on intermediate results, and inspect problematic samples when model performance is anomalous. However, supporting such human-in-the-loop interactivity in existing TEEs requires exposing I/O boundaries, creating severe risks of leaking private training data to reconstruction attacks~\cite{zhu2019deep, geiping2020inverting, song2017machine}.

\item \textbf{Exploratory Data Analysis:} In sensitive domains like healthcare or finance, data processors rarely formulate complete analysis programs in advance~\cite{chen2025picachv,kimLaputaSecureData2025}. Instead, they rely on iterative workflows to dynamically compose code, tune aggregation pipelines, and inspect distribution summaries without leaking raw records.

\item \textbf{Debugging and Diagnosis of Confidential Applications:} Maintaining complex applications in CVMs encounters a strict observability wall. When production containers crash, traditional diagnostics (\eg, attaching GDB or inspecting core dumps) are explicitly prohibited~\cite{ConfidentialcontainersReleasesV0170md,consortiumConfidentialComputingLogging, HowDebugConfidential}. Consequently, data processors urgently need secure mechanisms to interactively inspect runtime states, such as stack traces and variables, without compromising sensitive assets.

\end{packeditemize}

While these domains differ, they share a fundamental security dilemma: the necessity for dynamic code injection, runtime state observability, and non-deterministic operations directly conflicts with the static measurement and strict data confidentiality guarantees of traditional TEEs.

\section{Motivation}
\label{sec:motivation}


\subsection{The Dilemma between Interactivity and Measurability}
\label{sec:dilemma}

TEEs rely on remote attestation~\cite{RATS, coco_website} to guarantee the integrity of their initial memory states. This mechanism binds system trust to static measurements of the code and data loaded at boot time. Consequently, this security model establishes a strict chain of trust, requiring verifiers to depend entirely on these pre-launch measurements to confirm that the environment has booted into a known-good state.

However, a fundamental dichotomy exists between this rigid security model and the development lifecycle of complex applications. In scenarios such as machine learning model fine-tuning and exploratory data analysis, workloads are inherently dynamic and iterative~\cite{peft, chen2025picachv}. Data processors require the capability to observe intermediate results in real time, dynamically adjust hyperparameters, or even inject temporary debugging code to diagnose anomalies. 

This creates a fundamental dilemma between interactivity and measurability. Interactivity inherently implies the continuous injection of non-deterministic, unmeasured entropy (\eg, human inputs, dynamic code) into the system. Measurability, conversely, demands a static, deterministic boundary to establish a cryptographic root of trust. The current landscape places users in a zero-sum situation: to obtain TEE measurable security, they must forgo interactivity; conversely, to enable interactivity, they are forced to breach the integrity boundary, rendering the environment unmeasurable. Therefore, the critical question arises: how can we reconcile these two seemingly opposing properties within a unified framework?


\subsection{System Model}
\label{sec:sysmo}
We consider a representative system architecture comprising four distinct entities:

\bheading{Data Owners:} The entities with high-value sensitive data. They provision data access to data processors for specific computational tasks under the strict mandate that data confidentiality is preserved and misuse is strictly prevented.

\bheading{Data Processors:} The entities responsible for authoring code, debugging programs, or executing analytical tasks. They demand interactive access to the computing environment (\eg, via SSH) to dynamically submit non-deterministic instructions. Crucially, they view their source code as proprietary intellectual property, refusing to disclose plaintext execution traces for external auditing. 

\bheading{Auditors:} The verifying entities responsible for ensuring that the data processor's runtime behaviors strictly adhere to the owner's prescribed security policies. They may act directly on behalf of the data owners or serve as independent third-party verifiers.

\bheading{Cloud Service Provider (CSP):} The entity providing the underlying confidential computing environments. 

\subsection{Threat Model}
\label{sec:thrmo}
Our study makes the following assumptions:

\bheading{CSP is not trusted by data owners and data processors:} Consistent with traditional confidential computing models, we treat the CSP as an external adversary that controls the software stack outside the TEE, such as the hypervisor and host OS. This adversary may attempt to observe or tamper with protected execution through host-side attacks such as memory dumping, bus snooping, or privileged introspection, thereby targeting both the data owner's sensitive assets and the data processor's proprietary code or workflow artifacts. However, it is not assumed to break the TEE's hardware isolation guarantees~\cite{intel_tdx, amd_sev}. 

\bheading{Data processors are not trusted by data owners:} Interactive confidential computing introduces a second adversarial principal: the data processors. Unlike traditional confidential computing, where the attested workload is typically fixed and the main adversary is the CSP, interactive workloads intentionally permit data processors to issue new post-launch actions through legitimate interfaces. This threat does not assume that the TEE is compromised. Rather, it reflects a limitation of static remote attestation: attestation authenticates the initial measured state, but cannot pre-approve the data processor's future interactions. A malicious data processor may therefore abuse authorized interactivity to trigger policy-violating behavior, directly exfiltrate sensitive data, or infer protected information indirectly~\cite{zhu2019deep, fredrikson2015model}.

\bheading{Auditors are not trusted by data processors:} Establishing runtime compliance necessitates an auditing mechanism, which introduces a fundamental trust dilemma. Data processors inherently distrust auditors---including data owners acting as verifiers. Data processors consider their interactive workflows (\eg, proprietary code structure, API usage patterns and workflow strategy) to be highly valuable intellectual property. If the auditing mechanism requires plaintext execution traces or source code disclosure, a malicious auditor could exploit this access to steal the data processor's trade secrets. Consequently, the data processor views the auditor as a potential threat to its workflow privacy.


\bheading{Security Goals:} We target the following aims:
\begin{packeditemize}
    \item \textbf{Confidentiality:} The system must ensure the confidentiality of the data owner's assets against both the data processors and the CSP as a fundamental security invariant, which must hold even under interactive execution. Additionally, the confidentiality of the data processor's intellectual property (\eg, source code or workflow traces) must be protected from the CSP and the data owners. 
    

    \item \textbf{Integrity:} The system must guarantee the authenticity and policy compliance of dynamic workloads at runtime. While enforcing low-level control-flow integrity or arbitrary program functional correctness is intractable for non-deterministic, human-in-the-loop interactions, our integrity goal ensures runtime compliance. Specifically, it guarantees that the execution trace genuinely reflects the operations processed within the hardware-isolated environment, and that these dynamic behaviors strictly adhere to the data owner's prescribed IFC policies.

    \item \textbf{Verifiability:} The system must provide cryptographic mechanisms to prove that all aforementioned security invariants were strictly maintained throughout the interaction. Unlike static remote attestation, which authenticates only the initial state, the system must generate cryptographic evidence allowing auditors to formally verify the runtime compliance. Crucially, the verification process itself must not undermine any confidentiality goals. \qz{rewrote.}

\end{packeditemize}

\subsection{Design Goals and Technical Challenges}
\label{sec:intcc}

In light of the aforementioned conflicts and threats, we propose the paradigm of interactive confidential computing. We define this paradigm as a system that permits untrusted subjects to execute dynamic, non-deterministic operations within the protected boundary of a TEE, while simultaneously guaranteeing data confidentiality and verifiable runtime compliance.

\bheading{The Necessity of Logical Partitioning.} To resolve the dilemma described in \secref{sec:dilemma}, we argue that a logical partitioning of the TCB is necessary for the following reasons.


First, human interaction is inherently unmeasurable. Unlike deterministic algorithms, human operators introduce external, unpredictable entropy into the system. If these behaviors occur directly within the TCB, the runtime memory state is continuously mutated by unmeasured instructions.

Second, measurability is the prerequisite for trust. The security model of a TEE is predicated on the capability to measure the TCB. If the TCB includes unmeasurable interactive logic, the entire TCB becomes unmeasurable, invalidating the chain of trust established by remote attestation.

Consequently, we posit that interactivity is incompatible with a monolithic TCB. Logical partitioning is not merely an architectural choice but a necessity. To maintain security, we must strip the unmeasurable interactive logic from the TCB and isolate it into an \textit{Interactive Controller}, while confining measurable sensitive operations within a \textit{Verifiable Runtime}. This separation constitutes the only viable approach to reconciling dynamic control with static integrity.



To realize interactive confidential computing, the system must simultaneously satisfy the following three properties, which introduce non-trivial technical challenges:

\bheading{Property 1: Transparent Compatibility.}
As a primary usability requirement, the system must enable data processors to use existing technology stacks without being aware of the underlying partitioned architecture, while keeping performance overhead within an acceptable range. 
\begin{packeditemize}
    \item \textit{Challenge 1: The Performance Challenge and Compatibility Issues of Partitioned Execution.} The logical partitioning established above necessitates isolating interactive logic from data states, breaking the assumption of a single address space~\cite{abadi2016tensorflow}. If massive data such as model weights were transferred via traditional RPC, the overhead of serialization and deserialization would lead to unacceptable interactive latency. Therefore, a critical engineering challenge lies in how to create an illusion for upper layer applications into perceiving data as local within an isolated environment, while simultaneously eliminating the performance bottlenecks associated with cross-domain data copying.
\end{packeditemize}

\bheading{Property 2: Interactivity.}
The system must empower data processors to orchestrate the computational workflow and inspect intermediate states (\eg, dynamically modifying code, viewing statistical results). However, this interactivity must not devolve into a conduit for data leakage.
\begin{packeditemize}
    \item \textit{Challenge 2: Reconciling Flexibility with Information Flow Security.} In an interactive environment, distinguishing benign debugging interactions from malicious secret exfiltration is exceptionally challenging. Traditional access control lists only govern static file access. While information flow control can track data propagation paths~\cite{sabelfeld2003language}, strict noninterference properties often render the system unusable. The challenge lies in constructing a fine-grained security lattice within the TEE to constrain unstructured interactive behaviors within security boundaries compliant with relaxed noninterference~\cite{li2005downgrading}.
\end{packeditemize}

\bheading{Property 3: Verifiable Execution and Compliance.}
The system must generate cryptographic evidence of its dynamic runtime behavior. We refer to this property as \textit{verifiable runtime compliance}: it empowers data processors to cryptographically verify that their interactive instructions were indeed executed within the attested TEE, while enabling data owners and auditors to verify that the resulting execution trace strictly adhered to the prescribed security policies.

\begin{packeditemize}
    \item \textit{Challenge 3: The Efficiency and Privacy Challenge of Runtime Verification.} Prior approaches such as Operation Execution Integrity (OEI)~\cite{sunOAT} assume fixed operations whose legal control paths and critical data can be identified in advance; they are therefore ill-suited to interactive confidential computing, where execution paths are workload-dependent and not exhaustively enumerable. Furthermore, finer-grained attestation or instruction-level tracing would incur prohibitive overhead~\cite{shaControlFlowAttestationConcepts2024, 299850}. Directly disclosing execution traces may reveal the data processor's proprietary code structure or workflow strategy. Consequently, a key challenge is how to prove that dynamic operation sequences have been executed and adhered to prescribed policies, without sacrificing performance or data processor's privacy.   
    
\end{packeditemize}

\subsection{Our Solution}
\label{sec:solution}

To tackle the aforementioned challenges, we propose the \sysname framework and design the following mechanisms. 

First, a proxy-based dispatch system preserves the native development experience under partitioned execution while keeping sensitive objects inside the Verifiable Runtime. Second, a lattice-based IFC mechanism tracks data sensitivity and permits cross-boundary outputs only through approved declassification functions. Third, a privacy-preserving verifiable execution mechanism binds dynamic runtime traces to the attested Runtime and proves policy compliance in zero-knowledge without exposing the data processor's workflow.

\section{Framework Design}
\label{sec:framework}

\begin{figure*}[t]
    \centering
    \includegraphics[width=0.93\textwidth, trim=0.2cm 1cm 0.2cm 1cm, clip]{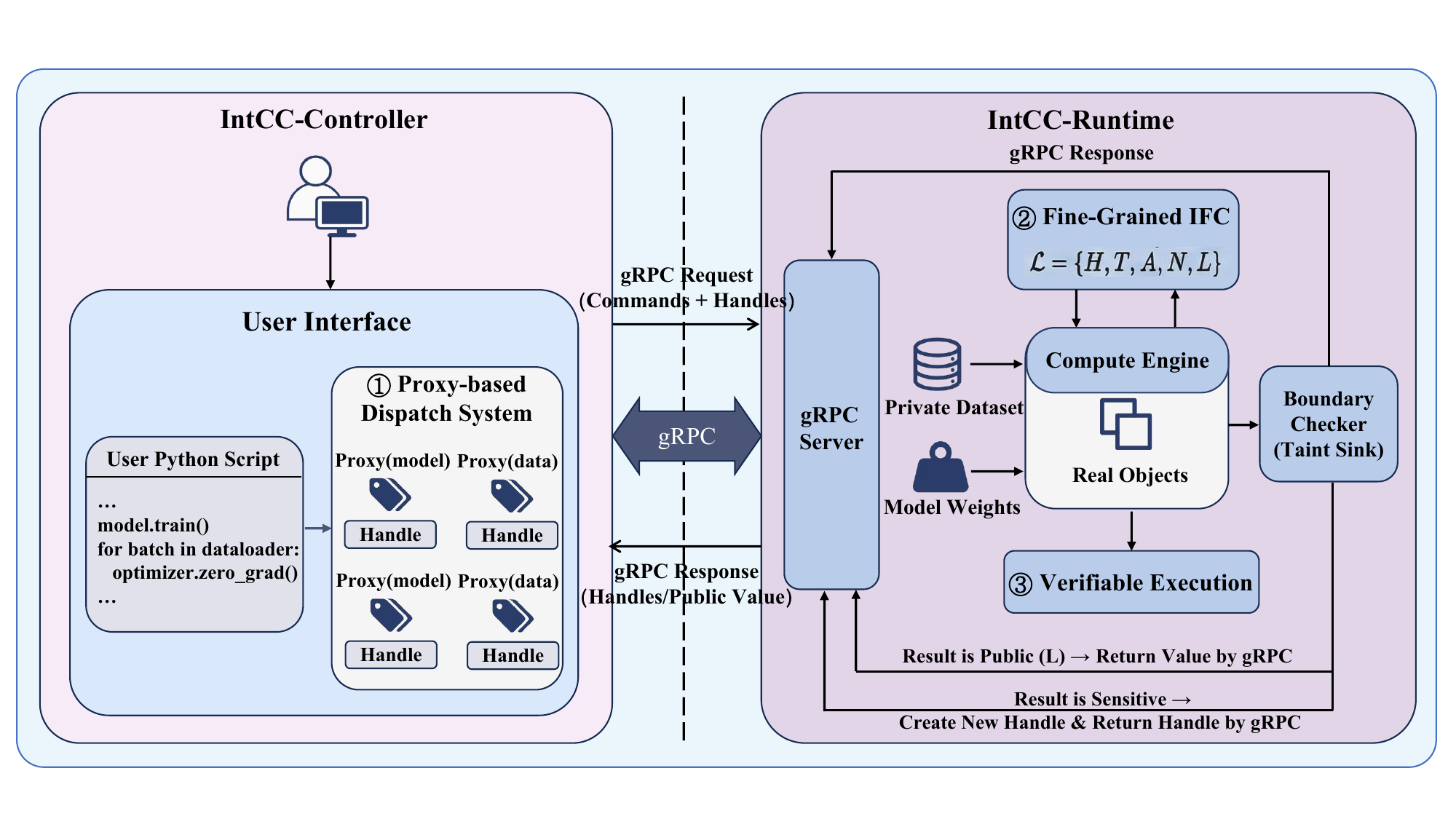} 
    \caption{High-level architecture of the framework.}
    \label{fig:architecture}
\end{figure*}

\subsection{Overall Architecture}

To resolve the tension between interactivity and measurability, \sysname materializes the logical partitioning strategy proposed in \secref{sec:dilemma}, separating an interactive control domain from a measured execution domain. As illustrated in \figref{fig:architecture}, the \textbf{IntCC-Controller}  hosts human-driven, post-launch logic and the proxy-based dispatch system. Conversely, the \textbf{IntCC-Runtime} hosts the compute engine operating on sensitive assets, securely governed by a fine-grained IFC mechanism and a verifiable execution module. While both partitions are physically protected from the untrusted host by the TEE, they assume distinct security roles: the Controller is excluded from the TCB for integrity, whereas the Runtime serves as the static root of trust for secure computation, strict policy enforcement, and runtime~evidence~generation. 

We formally define the roles and security properties of these two partitions as follows:

\begin{packeditemize}
        \item \textbf{IntCC-Controller:} This partition functions as the Interactive Controller, exposing standard development interfaces for dynamic post-launch interaction, including Python scripts, Jupyter kernels, and workload-specific control logic. It is structurally designed to host human-facing logic that cannot be fully pre-measured at boot time. Accordingly, although the confidentiality of this partition remains protected by the TEE against the CSP, it is strictly treated as outside the TCB for verifiable runtime compliance. From a policy enforcement perspective, the Controller is untrusted and may issue arbitrary interactive requests through the explicitly prescribed interfaces.
    
        \item \textbf{IntCC-Runtime:} This partition serves as the Verifiable Runtime and constitutes the system's measured root of trust for verifiable runtime compliance. It hosts the private data, execution backends, and the IFC monitor that mediates all dynamic interactions on sensitive assets. Initialized from an attested measured state, the Runtime is the foundational component whose identity is authenticated by remote attestation and whose audit evidence is later utilized to prove compliance with the prescribed IFC policies. This design ensures that all operations on sensitive data remain verifiable, fulfilling the integrity requirements.
   
\end{packeditemize}

As shown in the data flow in \figref{fig:architecture}, the IntCC-Controller cannot access any sensitive data directly. Instead, it interacts with the IntCC-Runtime through a proxy-based dispatch system (\secref{sec:proxy}). The Runtime enforces a fine-grained information flow control mechanism (\secref{sec:ifc}) to prevent data leakage and emits privacy-preserving runtime evidence via a verifiable execution mechanism (\secref{sec:lightweight}) to show that the trace was generated by the attested runtime and strictly satisfies the designated IFC constraints.   

Notably, this partitioned architecture supports diverse implementation strategies. It can be realized using two distinct SGX enclaves, isolated containers within a single CVM (\eg, leveraging CoCo~\cite{coco_website}), or a nested TEE configuration (\eg, executing the runtime within an Intel SGX~\cite{intel_sgx} enclave situated inside an Intel TDX~\cite{intel_tdx} trust domain).

\subsection{Proxy-based Dispatch System}
\label{sec:proxy}


To guarantee performance and preserve the native development experience under partitioned execution, \sysname incorporates a \textbf{proxy-based dispatch system}. By decoupling control flow from data flow, this system enables processors to manipulate remote assets using local syntax without exposing raw data. The design relies on two core mechanisms: \textit{shadow object abstraction} and an \textit{opaque reference protocol}. 

\subsubsection{Shadow Object Abstraction} In traditional remote execution, data is often serialized and transmitted to the client for processing. \sysname inverts this model by introducing \textit{shadow objects}---lightweight, stateless abstractions in the IntCC-Controller that act as semantic mirrors of the concrete entities residing within the IntCC-Runtime.

A shadow object acts as a proxy mirroring the public interface (attributes and methods) of the concrete class, encapsulating no physical data. When a data processor interacts with it, the object does not compute locally. Instead, it serves as a trap, intercepting invocations to preserve the native development experience while strictly separating states.

\subsubsection{Opaque Reference Protocol} 
\begin{figure*}[tbp]
    \centering
    \resizebox{0.95\textwidth}{!}{%
    \begin{tikzpicture}[
        node distance=1.2cm and 1.5cm,
        font=\sffamily\small,
        box/.style={draw, rounded corners, minimum width=2.8cm, minimum height=1.2cm, align=center, fill=white, line width=0.8pt},
        process/.style={box, fill=blue!5},
        data/.style={box, fill=yellow!10, shape=cylinder, shape border rotate=90, aspect=0.25, minimum height=1.5cm},
        proxy/.style={box, fill=green!5, dashed},
        arrow/.style={->, >=Stealth, thick, rounded corners},
        req_arrow/.style={->, >=LaTeX, ultra thick, color=blue!70!black},
        resp_arrow/.style={->, >=LaTeX, ultra thick, color=green!60!black, dashed}
    ]

    \node[process, align=left] (devScript) {\textbf{Script}\\ \texttt{y = function(x)}};
    
    \node[proxy, right=of devScript, text width=3.2cm] (proxyObj) {\textbf{Shadow Object}\\ (Interception)\\ Wraps: $h_x$};

    \node[proxy, below=2cm of devScript] (newProxy) {\textbf{New Shadow Object}\\ Wraps handle $h_y$};

    \node[process, right=6cm of proxyObj] (dispatcher) {\textbf{Dispatcher}};

    \node[process, below=1.8cm of dispatcher] (execEngine) {\textbf{Execution}\\ (Python Library)};

    \node[data, left=2cm of execEngine] (objStore) {\textbf{Object Store}\\ Registrate};

    \node[data, right=1.5cm of execEngine, minimum width=2cm] (realObj) {\textbf{Real Objects}};

    \begin{scope}[on background layer]
        \node[fit=(devScript)(proxyObj)(newProxy), draw, dashed, fill=gray!5, inner sep=0.5cm, label={[anchor=north west]north west:\textbf{IntCC-Controller} (Unmeasurable Domain)}] (controllerZone) {};
        
        \node[fit=(dispatcher)(realObj)(execEngine)(objStore), draw, dashed, fill=blue!5, inner sep=0.5cm, label={[anchor=north east]north east:\textbf{IntCC-Runtime} (Measurable Domain)}] (runtimeZone) {};
    \end{scope}

    \draw[arrow] (devScript) -- node[above, font=\footnotesize] {Invoke} (proxyObj);
    
    \draw[req_arrow] (proxyObj.east) -- node[above, font=\bfseries, align=center, yshift=0.1cm] {Dispatch\\ (Method + Handles)} node[below, font=\bfseries, color=red, align=center] {(NO RAW DATA)} (dispatcher.west);

    \draw[arrow] (dispatcher) -- (execEngine);
    \draw[arrow] (dispatcher.east) -| node[above, xshift=-0.5cm, font=\footnotesize] {Resolve} (realObj.north);
    \draw[arrow] (realObj.west) -- (execEngine.east);
    
    \draw[arrow] (execEngine.west) -- node[above, font=\footnotesize, align=center] {Execute} (objStore.east);

    \draw[resp_arrow] (objStore.west) -- node[above, font=\bfseries, color=green!40!black] {Propagate} node[below, font=\footnotesize, color=green!40!black] {(Return Handle $h_y$)} (newProxy.east);

    \draw[arrow] (newProxy.north) -- (devScript.south);

    \end{tikzpicture}
    }
    
    \caption{Workflow of the proxy-based dispatch system.}
    \label{fig:proxy_workflow_compact}
\end{figure*}
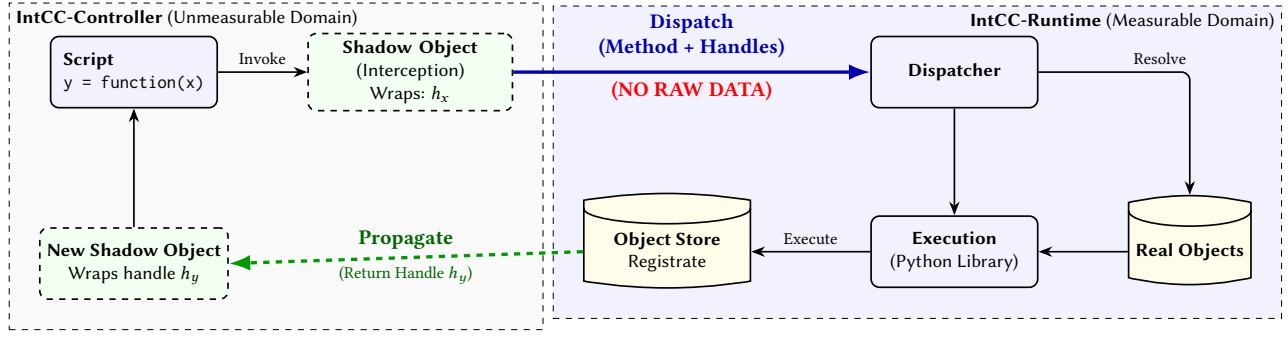

To secure the boundary, we employ a handle-based protocol that replaces direct memory access with opaque references. The workflow, as depicted in \figref{fig:proxy_workflow_compact}, proceeds in four logical phases:

\begin{packeditemize}
\item \textbf{Interception:} When an operation is triggered, the shadow object intercepts the data processor's call and marshals the method name alongside opaque handles---randomized identifiers that reveal zero information about the underlying data's content or memory layout---into a request.

\item \textbf{Dispatch:} The request is transmitted across the trust boundary to the IntCC-Runtime via a secure channel.

\item \textbf{Resolve and Execute:} The Dispatcher resolves the handle IDs to concrete memory addresses. The requested operation is then executed on the real data. 

\item \textbf{Register and Propagate:} The Runtime registers the execution result locally and propagates a unique Handle ID back to the Controller, where a new shadow object is instantiated to wrap it.

\item \textbf{Registration and Propagation:} Depending on the operation's semantics, the Runtime either updates the existing data in-place or registers a newly generated asset in its object store. When a new asset is produced, its unique Handle ID is propagated back to the Controller to instantiate a new shadow object, seamlessly sustaining the execution chain. 

\end{packeditemize}

Inspired by prior works like TrainCheck~\cite{traincheck} and DLBox~\cite{dlbox}, the proxy layer functions as both a state-isolation boundary and a dynamic policy enforcement point to regulate interactive operations. Specifically, the system inspects the data sensitivity level at this boundary: low-sensitivity data is permitted to be serialized and returned as concrete values, which enables data processors to seamlessly inspect low-sensitivity artifacts necessary for debugging and tuning. 

\subsection{Security Lattice Based Information Flow Control}
\label{sec:ifc}

To reconcile interactive flexibility with data confidentiality, \sysname employs a fine-grained IFC mechanism~\cite{denningifc, chen2025picachv}. This model formally defines data sensitivity levels within the IntCC-Runtime and strictly constrains information flow and declassification pathways during computation.

\subsubsection{Security Lattice and Taint Propagation}
We formalize the policy mechanism's foundation as a security lattice $\langle \mathcal{L},\sqsubseteq,\sqcup,\sqcap\rangle$. The label set $\mathcal{L}$ comprises five hierarchical levels dictating privacy obligations: \textbf{H} (High-Sensitivity) for strictly isolated raw data; \textbf{T} (Transformation) for data requiring masking or de-identification; \textbf{A} (Aggregation) for intermediate results needing batch averaging to prevent reconstruction; \textbf{N} (Noise Injection) for aggregated data mandating differentially-private noise~\cite{dwork2006differential} against differencing attacks; and \textbf{L} (Low-Sensitivity) for public artifacts cleared for export. These logically form a sequential privacy-preserving pipeline governed by a strict linear hierarchy (total order), satisfying $L \sqsubseteq N \sqsubseteq A \sqsubseteq T \sqsubseteq H$.

To enforce this, the mechanism utilizes $\mathcal{L}$ as dynamic taint markers, intercepting all IntCC-Runtime operations for automatic label propagation. We adopt the lattice Join ($\sqcup$) operation to handle information flow convergence, while the corresponding Meet ($\sqcap$) operation is reserved for policy composition. When an operation takes inputs with labels $l_1, \dots, l_k$, the output label is deterministically computed as the least upper bound of the inputs: $$l_{out} = \bigsqcup_{i=1}^{k} l_i$$ This design rigorously adheres to the noninterference principle~\cite{goguen1982security}, ensuring information only flows from lower to higher sensitivity levels, thereby preventing the inadvertent downgrading of highly sensitive assets during execution.

\subsubsection{Explicit Declassification and Boundary Audits}
While strict propagation prohibits unauthorized flow from high to low levels, practical usability dictates the need for controlled sensitivity reduction. We address this by replacing implicit downgrading with explicit declassification, executed via a trusted, audited exception process~\cite{sabelfeld2009declassification}.

\bheading{Explicit Declassification.} \sysname maintains a centrally managed registry that maps pre-audited, trusted sanitization functions to specific input/output label transition rules: \[f_{trusted}: l_{in} \to l_{out} \quad (\text{where } l_{out} \sqsubset l_{in})\] For instance, the registry may define: 
\begin{packeditemize}
\item $f_{dp}: N \to L$ (a differential privacy function that discharges the outstanding noise-injection obligation).
\item $f_{anonymize}: T \to L$ (a text anonymization function that discharges the outstanding transformation obligation).
\end{packeditemize}

The mechanism permits the output to bypass the default Join ($\sqcup$) propagation rule and acquire a lower sensitivity label only when the Runtime invokes these whitelisted functions. 

\bheading{Boundary Audits.} The system's security perimeter is enforced at the taint sink---specifically, the gRPC server response point where data returns to the IntCC-Controller. Crucially, to guarantee absolute confinement, \sysname strictly prohibits all other potential sinks at the container level (\eg, outbound network connections and arbitrary host file I/O are utterly disabled). Consequently, the gRPC interface serves as the singular, heavily audited exit point. The policy mechanism inspects all data attempting to cross this boundary: only data labeled $L$ is permitted to be serialized and returned as concrete values. For data at all other sensitivity levels, the system returns only opaque handles. 

\subsection{Privacy-Preserving Verifiable Execution Mechanism}
\label{sec:lightweight}

To guarantee verifiable runtime compliance, \sysname extends static remote attestation with runtime evidence. While remote attestation authenticates the IntCC-Runtime's launch state, the verifiable execution mechanism binds post-launch dynamic behavior to this attested runtime, proving policy compliance without exposing plaintext traces.

\subsubsection{TEE-Bound Trace Commitment}


The Runtime records each dynamic interaction as a canonical trace entry containing the operation name, operation class, input and output labels, whether an approved declassification function is invoked, whether the result is exported across the boundary, and whether the operation is allowed or denied. Let $\tau=(e_1,\ldots,e_n)$ denote the execution trace. The Runtime computes $C=\mathsf{SHA256}(\mathsf{ser}(\tau))$, and embeds $C$ into the hardware attestation report. This binds the trace commitment to a genuine measured Runtime instance, preventing an untrusted data processors or CSP from substituting a different trace, while keeping the plaintext trace private.

\subsubsection{Zero-Knowledge Proof and Verification}



Revealing $\tau$ to data owners or auditors would expose the data processor's proprietary workflow, including API usage patterns, model-tuning strategies, and intermediate control decisions. \sysname therefore treats $\tau$ as a private witness and proves only that it is both committed by $C$ and policy-compliant. Concretely, the Runtime generates a zero-knowledge proof $\pi_{zk}$ for the relation:
\begin{equation*}
    \mathcal{R}(C, \tau) = 1 \iff C = \mathsf{SHA256}(\mathsf{ser}(\tau)) \, \land \, \mathsf{CheckIFC}(\tau) = 1,
\end{equation*}
where $\mathsf{CheckIFC}$ verifies the IFC rules in \secref{sec:ifc}: ordinary operations must follow join-based taint propagation, declassification may occur only through approved transitions, and only $L$-labeled values may be exported. The public statement contains $C$ and the policy result, while the sequence of operations remains hidden.

Verification has two steps. First, auditors verify the hardware attestation report to check that $C$ was produced by an IntCC-Runtime with the expected static measurement. Second, they verify $\pi_{zk}$ against $C$ and the agreed policy circuit. A successful proof establishes that the committed trace satisfies the IFC policy without exposing the trace itself. Conversely, data processors can hash their local plaintext trace and compare it with the attested commitment $C$ to confirm trace authenticity and detect CSP-side omission or substitution. Thus, the mechanism gives data owners policy-compliance evidence and data processors trace-integrity evidence, without forcing either party to reveal private assets.

\subsubsection{Verification Scope}

The guarantee provided by this mechanism is intentionally narrower than full-program correctness. It does not prove that an arbitrary interactive workload implemented every unstated data processor's intent, nor does it certify each low-level computation step in isolation. Such step-by-step proof is unnecessary for the threat model because remote attestation already establishes that the execution trace is generated by the genuine attested Runtime and faithfully reflects its actual execution. The remaining task is therefore not to re-prove each low-level execution step, but to verify that the recorded execution complied with the designated policies. To this end, \sysname provides \emph{verifiable runtime compliance}. Under this guarantee, data processors can verify that their interactions were processed within the attested Runtime, while data owners or auditors can verify that the recorded execution remained policy compliant.
\section{Implementation}

We implemented the core framework of \sysname in Python, comprising approximately 3,200 Lines of Code (LoC), and deploy it on AMD SEV-SNP~\cite{amd_sev} using Confidential Containers~\cite{coco_website}. We chose Python to natively support dynamic computational graphs, runtime reflection, and dynamic object instantiation, which are prevalent in modern AI and data analytics stacks like PyTorch~\cite{pytorch} and Dask~\cite{dask}. Furthermore, this general design minimizes system extension effort, requiring no more than \textbf{200 LoC} to support new complex orchestration frameworks (\eg, LLaMA-Factory~\cite{llamafactory}, DuckDB~\cite{duckdb}).


\bheading{Implementation of Proxy-based Dispatch System.} We realize the shadow object abstraction through dynamic interception of method invocation, attribute access, and object construction on the controller side. Each intercepted operation is marshaled into a Protobuf request containing the operation name, opaque runtime-object handles, and any public literal arguments. On the IntCC-Runtime, a dispatcher resolves handles through a thread-safe object registry, invokes the backend operation, and registers returned objects under fresh handles. To support dynamic Python libraries that rely on runtime inspection and modification, we further implement a remote reflection path that forwards metadata queries to the Runtime and synthesizes matching controller-side stubs on demand, thereby preserving native Python development without source changes.

\bheading{Implementation of IFC mechanism.} We implement IFC as a dynamic enforcement layer in the Runtime dispatcher. Each runtime object is stored together with label metadata, and every invocation is checked against a policy map. Concretely, the Runtime deserializes the receiver and arguments, computes an effective input label by joining their tags, resolves the corresponding policy rule, and validates that the invocation is admissible under that rule. The Runtime then either denies the operation or executes it and derives the output label according to the resolved policy. In the current policy map, all non-declassification operators use the default join rule, while trusted operators implement explicit declassification transitions such as anonymization and DP release. We integrate IBM Diffprivlib~\cite{diffprivlib} for differentially private numerical release and Microsoft Presidio with spaCy~\cite{Honnibal_spaCy_Industrial-strength_Natural_2020} for text anonymization. A serialization guard at the response boundary then enforces the final export rule: only objects labeled $L$ may leave the Runtime as concrete values; higher-labeled outputs remain inside the Runtime and are returned as opaque handles or rejected.

\bheading{Implementation of Verifiable Execution.} We implement runtime evidence generation as a service inside the Runtime. Whenever the IFC monitor processes a dynamic operation, it appends the corresponding \texttt{IFCTraceEntry} to the in-memory trace. At the end of task execution, the Runtime canonicalizes the trace, computes a SHA-256 trace commitment $C$, and embeds $C$ into the attestation report. For privacy-preserving verification, we implement a zero-knowledge proof backend based on SP1~\cite{sp1_github}: the prover takes the plaintext execution trace as private input, re-encodes each entry into the compact witness format expected by the circuit, and generates a proof that (1) the witness trace hashes to the attested commitment $C$, and (2) every trace entry satisfies the IFC rules; the verifier then checks the resulting proof against the public output and the verification key deterministically derived from the guest ELF.

Finally, we instantiate \sysname across two representative domains. For interactive fine-tuning, \sysname supports native PyTorch scripts~\cite{pytorch}, PEFT-based adapter tuning~\cite{peft}, and LLaMA-Factory~\cite{llamafactory}. For analytical workloads, it supports Dask~\cite{dask}, DuckDB~\cite{duckdb}, and Pandas~\cite{pandas} over the TPC-H benchmark~\cite{tpch}. Together, these instantiations show that the implementation is not tied to any single stack, but to a reusable Runtime for heterogeneous interactive workloads. Crucially, framework extension overhead is minimal: supporting any of these complex ecosystems requires modifying fewer than 200 LoC within the core orchestration logic.
\section{Evaluation}

In order to clearly demonstrate the practical impact of \sysname, we design an experiment to empirically evaluate the performance overhead and security enhancement of \sysname. We aim to answer the following research questions:
\begin{packeditemize}
    \item \textbf{Performance Efficiency:} Does the logical partitioning architecture introduce prohibitive latency compared to standard, monolithic confidential execution?  (\secref{sec:performance})

    \item \textbf{Security Guarantees:} Can the system reliably prevent data exfiltration and enforce verifiable runtime compliance under adversarial conditions? (\secref{sec:security})
\end{packeditemize}

\bheading{Experimental Setup.} All experiments were conducted on a server equipped with an AMD EPYC 9334 processor, employing AMD SEV-SNP~\cite{amd_sev} for hardware-level memory encryption. An NVIDIA H100 GPU was attached via PCIe passthrough. The software stack was hosted on Ubuntu 24.04.3 LTS, utilizing Confidential Containers to orchestrate the confidential execution sandbox~\cite{coco_website}.

\subsection{Performance Evaluation}
\label{sec:performance}

We focus on quantifying the specific overhead imposed by the partitioned architecture of \sysname. To do this rigorously, we define two distinct architectural configurations:

\begin{packeditemize}
    \item \textbf{Baseline (Monolithic TEE):} This configuration executes the target computational workflows within a single, monolithic Confidential Container. This represents the current state-of-the-art for static confidential workloads, where the training script, data, and computation engine all reside in the same trusted domain with local access.
    \item \textbf{\sysname (Partitioned TEE):} This setup executes the same workflows using the proposed partition architecture. The interactive logic runs in the Controller, while sensitive data and computation reside in the Runtime. All interactions are mediated via the proxy-based dispatch system. 
\end{packeditemize}

We define overhead as the relative increase in end-to-end latency induced by the \sysname partitioned architecture. By benchmarking \sysname against a monolithic confidential baseline rather than a native, untrusted host process, we effectively factor out standard TEE virtualization costs. This methodology strictly isolates the performance overhead of interactivity: the latency introduced by gRPC communication, IFC enforcement, zero-knowledge proof generation, and the proxy-based dispatch system.


\begin{figure*}[t]
    \centering
    \includegraphics[width=\linewidth]{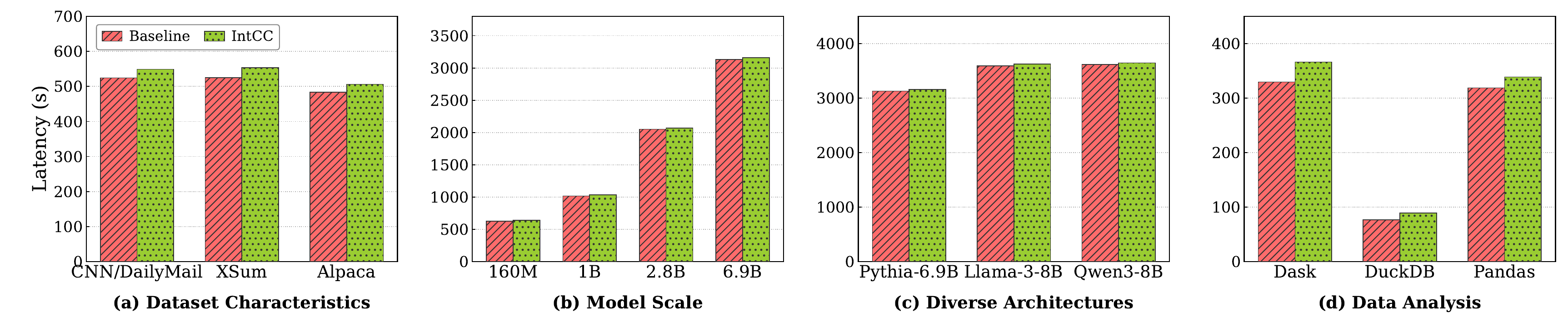} 
    \caption{End-to-end performance analysis.}
    \label{fig:perf_analysis}
\end{figure*}

To comprehensively evaluate the performance overhead introduced by \sysname, we structured the experiments across two primary categories: Fine-tuning and data analysis.

\subsubsection{LLM Fine-tuning}

\tabref{tab:exp_setup} details the experiment configurations, listing the evaluation dimensions, datasets, and model architectures employed for each dimension. Unless otherwise specified, all tasks utilized Hugging Face PEFT~\cite{peft} to replicate mainstream industrial usage patterns.

\begin{table}[htbp]
\centering
\footnotesize
\caption{Details of datasets and models used in evaluation.}
\label{tab:exp_setup}
\begin{tabularx}{\linewidth}{@{} >{\raggedright\arraybackslash}X >{\centering\arraybackslash}X >{\centering\arraybackslash}X c @{}}
\toprule
\textbf{Dimension} & \textbf{Dataset} & \textbf{Model} & \textbf{Size} \\
\midrule
Dataset Characteristics & 
\makecell[tc]{CNN/DailyMail \\ XSum \\ Alpaca} & 
Pythia & 
6.9B \\
\midrule
Model Scale & 
CNN/DailyMail & 
Pythia Family & 
\makecell[tc]{160M, 1B, \\ 2.8B, 6.9B} \\
\midrule
Diverse Architectures & 
CNN/DailyMail & 
\makecell[tc]{Pythia \\ Llama3 \\ Qwen3} &
\makecell[tc]{6.9B \\ 8B \\ 8B} \\
\bottomrule
\end{tabularx}
\end{table}

We first evaluated the end-to-end performance under diverse dataset characteristics. We selected three widely-used benchmark datasets—CNN/Daily\-Mail~\cite{see-etal-2017-get}, XSum~\cite{Narayan2018DontGM}, and Alpaca~\cite{peng2023gpt4llm}—to represent diverse sequence lengths and workload patterns (evaluating 1,000 instances each). Specifically, the processing time incurred marginal overheads of 4.50\% for CNN/DailyMail, 4.55\% for XSum, and 4.53\% for Alpaca.


To evaluate the impact of workload complexity, we performed a scaling analysis on 10,000 samples using the Pythia family~\cite{biderman2023pythia}, ranging from 160M to 6.9B. \figref{fig:perf_analysis}(b) demonstrates a clear \textit{overhead amortization} effect: as the model size increases, the relative performance overhead consistently decreases. This phenomenon occurs because the fixed system costs are effectively amortized over the dominating GPU computation time required by larger models. Specifically, the lightweight Pythia-160M incurred a 2.5\% overhead, which dropped to 1.7\% for Pythia-1B, and further to 0.9\% for Pythia-2.8B. For the largest model, Pythia-6.9B, the overhead stabilized at a mere 0.9\%. 


To verify that this performance consistency extends beyond a single architecture, we further evaluated \sysname across other prominent foundation models, including Llama-3-8B~\cite{llama3modelcard} and Qwen3-8B~\cite{qwen3technicalreport}. For each model, we fine-tuned 10,000 randomly selected samples. \figref{fig:perf_analysis}(c) demonstrates that \sysname maintains high efficiency regardless of the specific foundation model. For Llama-3-8B, \sysname was only 0.81\% slower than the native baseline, while the overhead for Qwen3-8B was strictly controlled at 0.75\%. 


In summary, the evaluation across these three categories comprehensively validates the efficiency and robustness of \sysname in fine-tuning applications. Furthermore, the \textit{overhead amortization} effect observed with increasing model scales confirms that fixed CPU-bound costs such as gRPC and security enforcement are effectively amortized in compute-intensive fine-tuning scenarios. Ultimately, the consistent performance across complex, modern model families like Llama-3~\cite{llama3modelcard} and Qwen3~\cite{qwen3technicalreport} highlights the generality of \sysname, seamlessly supporting large-scale LLM fine-tuning tasks without requiring manual adaptation or incurring additional translation penalties.

\subsubsection{Data Analysis}

To evaluate the efficacy of \sysname on data analysis workloads, we benchmarked \sysname using a TPC-H dataset generated at scale factor 10 (SF=10)~\cite{tpch}. 

We compared three mainstream data analysis frameworks that have been successfully integrated into \sysname: Dask~\cite{dask} DuckDB~\cite{duckdb}, and Pandas~\cite{pandas}. As illustrated in \figref{fig:perf_analysis}(d), the frameworks exhibit distinct performance behaviors under \sysname architecture.Specifically, Pandas incurred the lowest end-to-end overhead of 6.32\%, followed by Dask with a moderate overhead of 11.16\%, while DuckDB exhibited the highest percentage overhead at 16.56\%.


In summary, maintaining an end-to-end overhead strictly below 17\% represents a highly acceptable and competitive trade-off for practical deployments. More importantly, the demonstrated ability to seamlessly support architecturally diverse and complex frameworks highlights its strong practical applicability and broad generality of \sysname, proving its capacity to securely orchestrate a wide spectrum of real-world computing paradigms.


\subsubsection{Micro-benchmark}







\begin{table}[htbp]
\centering
\caption{Overhead breakdown across different workloads.}
\label{tab:overhead_breakdown}
\footnotesize
\setlength{\tabcolsep}{3pt} 
\begin{tabular}{lccccc}
\toprule
\textbf{Workload} & \textbf{Total (s)} & \textbf{gRPC} & \textbf{IFC} & \textbf{ZK Proof} & \textbf{Proxy} \\
\midrule
Dask     & 36.78 & 0.16 (0.4\%) & 0.00 (0.0\%) & 8.57 (23.3\%) & 28.06 (76.3\%) \\
DuckDB   & 12.68 & 0.16 (1.3\%) & 0.00 (0.0\%) & 8.81 (69.5\%) &  3.71 (29.3\%) \\
Pandas   & 20.15 & 0.20 (1.0\%) & 0.00 (0.0\%) & 8.72 (43.3\%) & 11.23 (55.8\%) \\
Finetune & 27.70 & 0.06 (0.2\%) & 0.07 (0.2\%) & 9.27 (33.5\%) & 18.30 (66.1\%) \\
\bottomrule
\end{tabular}
\end{table}


To further investigate the composition of system overhead, we conducted a comprehensive micro-benchmark across both fine-tuning and data analysis workloads, as detailed in \tabref{tab:overhead_breakdown}. We selected the Pythia-6.9B task as the representative of LLM fine-tuning. The breakdown reveals that across all categories, gRPC network latency and IFC enforcement contribute negligibly to the total overhead. With the exception of DuckDB, the \textit{Proxy System Overhead}, stemming from the proxy-based dispatch system, constitutes the largest percentage of the latency: specifically, it accounts for 76.3\% in Dask, 66.1\% in Pythia-6.9B fine-tuning, and 55.8\% in Pandas.

We attribute this phenomenon primarily to the proxy-based dispatch system execution environment. Data-intensive operations are inherently sensitive to the runtime state; when executing within a gRPC worker thread rather than a clean local main thread, concurrent live RPC dispatching and poller activities significantly elongate the wall-clock coordination path of the underlying task schedulers. Conversely, DuckDB exhibits a unique distribution where zero-knowledge proving accounts for the majority of its overhead (69.5\%). This is due to the \textit{overhead amortization} effect: because DuckDB's absolute execution penalty is exceptionally low, the cryptographic costs become proportionally dominant. Overall, these results demonstrate that although proxy-based dispatching constitutes the primary source of latency, its absolute impact remains highly practical across architecturally diverse workloads.

\subsection{Security Evaluation}
\label{sec:security}

In this section, we evaluate the security effectiveness of \sysname in the context of LLM fine-tuning. The evaluation is twofold: we first validate the resilience of \sysname against adversarial interactions, and subsequently quantify the efficacy of its declassification mechanisms in realistic \yz{debugging?}\qz{removed.} scenarios.

\subsubsection{Defense against Adversarial Interactions} 
We evaluate the system's resilience against three primary threat vectors: explicit data theft by data processors, implicit data reconstruction by data processors, and execution pipeline tampering by the untrusted cloud provider.
\yz{one sentence to summarize what kind of adverarial interactions, what are the relationship between the two bheadings.}\qz{added.}

\bheading{Defense against Direct Data Theft.} Malicious data processors may attempt to exfiltrate sensitive assets via I/O channels, including standard output streams, file system updates, or network transmission. To evaluate this, we simulated a malicious data processor and injected interactive commands attempting to invoke prohibited I/O operations on the sensitive data, such as printing plaintext tensors to the terminal, writing variables to \texttt{.pth} files, and sending outbound HTTP requests. \sysname successfully neutralized $100\%$ of these attempts. The defense is attributed to the IFC enforcement, which correctly identified the target objects as High-sensitivity and intercepted the serialization requests at the system boundary. Consequently, the system returned only opaque handles to the controller, preventing any bit-stream transmission to the external environment.

\bheading{Defense against Data Reconstruction Attacks.} Attackers may attempt to reconstruct private training data by performing gradient inversion~\cite{geiping2020inverting}. Specifically, the adversary captures the gradient $\nabla W$ generated during a training step and iteratively optimizes a dummy input until its resultant gradient matches the intercepted $\nabla W$, thereby recovering the original input. \yz{this attack step can be polished, not very accurate and clear.}\qz{rewrote.} We simulated this threat by crafting requests to export exact gradient tensors during the backward pass \yz{this is also unclear.}\qz{rewrote.}. \sysname successfully thwarted these attempts. The defense is attributed to the strict enforcement of DP policies on intermediate computation states. The exposed gradients injected by noise lack the precision required for the inversion optimization to converge, rendering reconstruction of the private training samples \yz{reconstruction of what}\qz{added.} infeasible.

\bheading{Defense against Pipeline Tampering.} A lazy or malicious cloud provider may attempt to compromise the integrity of the training pipeline, for example, by skipping computations to save GPU resources. We simulated this threat by configuring a malicious gRPC proxy to intentionally drop execution requests for the \texttt{optimizer.step()} call during training loop iterations.\yz{attack steps are not clear, how to skip training steps?}\qz{added.} The evaluation shows that the omission of these specific steps was successfully detected upon audit. This defense capability is attributed to the verifiable execution mechanism. Any deviation from the canonical execution path generates a runtime hash that mismatches the cryptographic commitment anchored in the attestation report, immediately flagging the pipeline tampering.


\subsubsection{Efficacy of PII Sanitization Engine}
\label{sec:pii}

To validate the utility of safe data previews, we evaluated the PII sanitization engine in identifying and masking sensitive entities. We utilized a subset of the OntoNotes 5.0 dataset~\cite{hovy-etal-2006-ontonotes} (5,000 samples). \tabref{tab:pii_efficacy} presents the evaluation results. The engine showed high efficacy across critical entity categories, achieving an overall F1-Score of 0.9510. These results indicate that the system effectively minimizes the attack surface during data previews while still allowing data processors to comprehend structural features and debug their code. 

\begin{table}[t]
\centering
\footnotesize
\caption{Efficacy of the PII anonymization.}
\label{tab:pii_efficacy}
\begin{tabular}{lccc}
\toprule
\textbf{Entity Type} & \textbf{Precision} & \textbf{Recall} & \textbf{F1-Score} \\
\midrule
EVENT               & 0.8000             & 0.9836          & 0.8824            \\
FAC               & 0.9643             & 0.9469          & 0.9555            \\
GPE         & 0.9620             & 0.9701 & 0.9660            \\
PRODUCT                 & 0.9371             & 0.9982          & 0.9667            \\
TIME                  & 1.0000             & 0.8579          & 0.9235            \\
\midrule
\textbf{Overall}     & 0.9465             & 0.9556          & \textbf{0.9510}   \\
\bottomrule
\end{tabular}
\end{table}

 \begin{figure}[t]
    \centering
    \includegraphics[width=0.85\columnwidth]{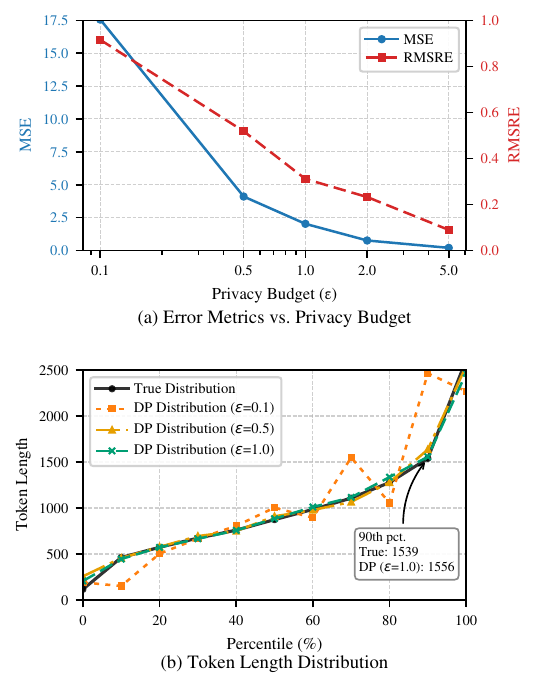}
    \caption{Evaluation of the privacy-utility trade-off.}
    \label{fig:safety_utility}
  \end{figure}


\subsubsection{Privacy-Utility Trade-off in Differential Privacy} We further evaluated the inherent trade-off between privacy guarantees and debugging utility when applying DP mechanisms. This evaluation assesses whether the sanitized statistics provided by \sysname are sufficient for data processors to make correct decisions regarding hyperparameter tuning and training stability.

\bheading{Utility Quantification.} As a baseline heuristic for statistical fidelity, we first observed the deviation of noisy gradients from ground-truth values during a Pythia-160M fine-tuning task (using a subset of 2,000 CNN/DailyMail~\cite{see-etal-2017-get} samples, trained over 10 epochs with a batch size of 16, and a chunk size of 100 for step aggregation). As illustrated in \figref{fig:safety_utility} (a), the Mean Squared Error (MSE) and Root Mean Squared Relative Error (RMSRE) decline rapidly as the privacy budget ($\epsilon$) increases. At a moderate budget of $\epsilon=2.0$, the MSE stabilizes at 0.7535. While MSE serves primarily as a macro-level indicator, this bounded error margin suggests that the sanitized gradients successfully preserve the overall distributional trends of the original data. This provides data processors with the necessary directional signals to monitor for training anomalies (\eg, gradient vanishing or exploding).

\bheading{Case Study: Assisted Hyperparameter Configuration.} To validate practical utility, we conducted a case study on configuring the \texttt{max\_length} hyperparameter, a critical setting that balances computational efficiency against information loss due to truncation\yz{truncation.}. We computed the quantile distribution of token lengths for the private dataset (a subset of 2,000 samples from the CNN/DailyMail dataset) via the DP mechanism. \figref{fig:safety_utility} (b) compares the ground-truth distribution with the DP-protected output. The results show that the noisy 90th percentile value (1556) is statistically equivalent to the ground truth (1539). A data processor relying solely on the sanitized view would derive the same optimal configuration (\eg, setting \texttt{max\_length=2048}) as one with direct data access. This confirms that \sysname successfully preserves the utility required for interactive development workflows while maintaining strict privacy guarantees.

\section{Discussion}

We reflect on the limitations of \sysname and outline future directions. It is worth noting that the framework remains agnostic to specific implementations, enabling components to evolve with technological advancements.


\bheading{Overhead of Zero-Knowledge Proving.} A practical limitation of current design is the cost of zero-knowledge proving: while it avoids exposing plaintext audit traces, proof generation is not lightweight enough for interactive confidential workloads. In particular, prover latency and resource consumption can become a bottleneck when audit frequency grows or when the system scales to larger distributed deployments. An important next step is to reduce proof-generation overhead through more efficient SNARK/STARK backends~\cite{Gabizon2019PLONKPO,BenSasson2018ScalableTA} or hardware acceleration~\cite{Lu2023cuZKAZ}.


\bheading{Programmable Security Policies.} The effectiveness of \sysname relies heavily on the expressiveness and correctness of the security policies defined by data owners. Currently, declassification rules are statically registered. We argue that data processors should be empowered to upload custom declassification logic encapsulated in sandboxed formats like WebAssembly. The core challenge then shifts to automated verification. Future iterations of \sysname could adopt a Proof-Carrying Code (PCC)~\cite{ProofcarryingCode} model, where data processors submit code alongside a formal proof demonstrating adherence to security policies. The IntCC-Runtime would be responsible for validating the proof~\cite{agora}.


\bheading{Limitations against Covert Channels.} While \sysname effectively constrains explicit information flows and partial implicit flows, these guarantees do not extend to covert channels (\eg, timing, termination, scheduling, or resource contention)~\cite{sabelfeld2003language}. Achieving progress-sensitive or timing-sensitive noninterference requires supplementary interventions, such as rigorous program analysis, deterministic padd\-ing, or secure multi-execution~\cite{moore2012precise,kashyap2011timing,devriese2010noninterference}. Consequently, \sysname provides robust semantic-level confinement, while comprehensive defense against covert channels remains an orthogonal challenge requiring complementary mechanisms.


\section{Related Work}
\label{sec:related_work}

\bheading{Confining Untrusted Software in TEEs.} NestedSGX~\cite{NestedSGX} and vSGX~\cite{vSGX} aim to minimize the TCB via guest-level \textit{privilege separation}, which necessitates extensive modifications to the hypervisor or guest OS. In contrast, \sysname functions as an application-level logical partition, rendering it orthogonal to, and deployable atop, these system-level designs.

Another line of research enforces constraints at the \textit{language or framework level}. For instance, PoBF~\cite{pobf} enforces security objectives through rigorous compile-time static analysis in Rust. Although this approach ensures safe execution paths for deterministic tasks, it relies on the assumption of a predictable computational graph, making it ill-suited for the exploratory, dynamic nature of data science (\eg, Python programs). Unlike these strict static constraints, \sysname targets the flexibility required by interactive analysis.

\bheading{Privacy-Preserving Model Tuning and Data Analytics.} \textit{Privacy-Preserving fine-tuning systems} represent one important line of work. Atlas~\cite{atlas} focuses on auditing the machine-learning lifecycle but does not address secure interactivity against untrusted data processors. DLBox~\cite{dlbox} pioneers partition and proxy mechanism for sensitive training data, but its scope is restricted to model training workloads, lacking the versatility of a general-purpose, interactive confidential computing interface.

\textit{Privacy-preserving data analytics systems} like Laputa~\cite{kimLaputaSecureData2025} secure Apache Spark through isolated execution and fine-grained policy checking over physical plans. Similarly, PICACHV~\cite{chen2025picachv} formally verifies data-use policy enforcement over relational-algebra query plans, demonstrating this approach in Polars. While these systems provide workload-specific guarantees for analytics engines, \sysname offers a workload-agnostic framework capable of supporting LLM fine-tuning, data analysis, and other interactive tasks.

\section{Conclusion}


In this paper, we introduced the interactive confidential computing paradigm to reconcile static TEE integrity with dynamic workflows. To operationalize this paradigm, we designed \sysname. By logically decoupling control from execution, enforcing lattice-based information flow control, and guaranteeing verifiable runtime compliance, \sysname securely enables human-in-the-loop operations. Evaluations confirmed \sysname supports complex tasks like LLM fine-tuning and data analysis with practical overhead.




\bibliographystyle{ACM-Reference-Format}
\bibliography{references} 


\end{document}